# Finite-difference frequency-domain study of subwavelength lensing in left-handed materials

X. Wang,[a,b] Stephen K. Gray,[b] George C. Schatz[a]

[a]*Department of Chemistry, Northwestern University, Evanston, IL 60208*

[b]*Chemistry Division and Center for Nanoscale Materials, Argonne National Laboratory,*

*Argonne, IL 60439*

**Abstract:** We show that the finite-difference frequency-domain method is well-suited to study subwavelength lensing effects in left-handed materials (LHM's) and related problems. The method is efficient and works in the frequency domain, eliminating the need for specifying dispersion models for the permeability and permittivity as required by the popular finite-difference time-domain method. We show that "superlensing" in a LHM slab with refractive index $n$ = -1 can be approached by introducing an arbitrarily small loss term. We also study a thin silver slab, which can exhibit subwavelength imaging in the electrostatic limit.

## 1. Introduction

A left-handed material (LHM) has simultaneously negative permittivity, $\varepsilon$, and permeability, $\mu$, and refractive index $n = \varepsilon^{1/2} \mu^{1/2} < 0$. Proposed by Veselago [1] in 1968, LHM's are of much current interest [2, 3]. Pendry [4] argued theoretically that a LHM slab with $n = -1$, in air, amplifies evanescent waves as well as reverses the phase of propagating components, allowing reconstruction of subwavelength features of an object. Without material losses perfect resolution can be achieved, which has been termed "superlensing."

Many numerical simulations of LHM's, and related problems employ the finite-difference time-domain (FDTD) method [5], e.g. Refs. [6-8, 11, 12]. The FDTD approach is conceptually simple, involves relatively straightforward grid representations, and is a well-established modeling tool. However, negative $\varepsilon$ and $\mu$ values are cumbersome to implement in FDTD simulations. Instabilities arise with the most obvious approach of simply setting $\varepsilon$ and $\mu$ to be negative numbers. Dispersion relations for the frequency dependence of $\varepsilon$ and $\mu$ must instead be specified, and auxiliary differential equations or recursive methods, equivalent to the dispersion relations, are incorporated into the FDTD simulation [6-8]. Furthermore, to obtain a steady state in a reasonable amount of time, loss (Im[$\varepsilon$], Im[$\mu$] $\neq 0$) must be included in the dispersion relations [6, 7]. Because of the very sensitive dependence of subwavelength imaging on deviations of $\varepsilon$ and $\mu$ from very specific values [9], it is desirable to investigate other numerical approaches that allow a more facile specification of the material parameters.

We show here that the finite-difference frequency-domain method (FDFD) [13, 14], in which we can set $\varepsilon$ and $\mu$ to specific (complex) values, provides an efficient and good description of LHM subwavelength imaging. At the same time, it also retains much of the simplicity and convenience of the FDTD method. The approach still involves introduction of loss to maintain stability. However, we find that extremely small loss terms suffice. The two-dimensional imaging system we consider here is a planar LHM slab surrounded by vacuum or asymmetric media with different positive refractive indices, as shown in Fig. 1. We also consider the case of a silver slab.

Section 2 presents our implementation of the FDFD method. A lens given by a slab with $\varepsilon = \mu = -1$, surrounded by air, is studied in Sec. 3. Section 4 addresses a non-ideal but practical lens system corresponding to a thin silver slab in an environment similar to that studied experimentally in Ref. [16].

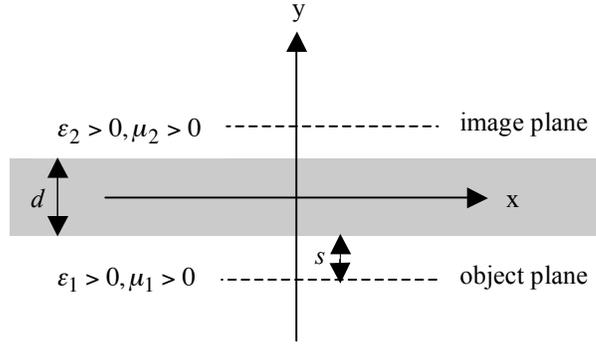

Fig. 1. Configuration of the two-dimensional imaging system. The shaded region is taken to be either a LHM or silver. For all the simulations presented, the calculation domain is 1000 nm x 100 nm.

## 2. Finite-difference frequency-domain (FDFD) method

The vector Helmholtz equations, in the absence of external currents, are

$$\nabla \times \frac{1}{\mu} \nabla \times \mathbf{E} - \omega^2 \varepsilon_0 \mu_0 \, \varepsilon \, \mathbf{E} = 0 \tag{1}$$

and

$$\nabla \times \frac{1}{\varepsilon} \nabla \times \mathbf{H} - \omega^2 \varepsilon_0 \mu_0 \, \mu \, \mathbf{H} = 0 \quad , \tag{2}$$

where **E**(*x,y,z*) and **H**(*x,y,z*) are the complex or phasor representations of the electric and magnetic fields associated with time dependence exp(-*iωt*). Here, $\omega = 2\pi c/\lambda$ is the frequency, and the relative permittivity and permeability vary in space: $\varepsilon = \varepsilon(x,y,z)$ and $\mu = \mu(x,y,z)$. FDFD methods are numerical approaches for solving Eq. (1) or (2) that incorporate many aspects of the FDTD method [5].

The problems of interest here can be addressed with a 2-D version that assumes invariance of *ε* and *μ* along *z*. In this case one can define two independent polarizations which each involve just three of the six possible electromagnetic field components: transverse magnetic (TM) polarization involving $E_x$, $E_y$, and $H_z$, and transverse electric (TE) polarization involving $H_x$, $H_y$, and $E_z$. We define transverse to be perpendicular to the plane of incidence, which is assumed to be the *x-y* plane. (Another convention is to define transverse to be perpendicular axis of invariance, *z*, which leads to an unfortunate reversal of the meanings of TE and TM.) The TM case involves just one magnetic field component, $H_z$. We can then focus on solving the corresponding Helmholtz equation for $H_z$ deduced from Eq. (2),

$$\frac{\partial}{\partial x}\frac{1}{\varepsilon}\frac{\partial H_z}{\partial x} + \frac{\partial}{\partial y}\frac{1}{\varepsilon}\frac{\partial H_z}{\partial y} + \omega^2 \varepsilon_0 \mu_0 \mu \, H_z = 0 \; . \tag{3}$$

The relation $\mathbf{E} = [i/(\omega\mu_0\mu)] \, \nabla \times \mathbf{H}$ is used to obtain $E_x$ and $E_y$ from $H_z$, when needed. Let $H_z = H_z^{in} + H_z^{sc}$, where $H_z^{in}$ is an arbitrary incident wave defined over the whole space. The linearity of the Helmholtz equation implies that

$$\begin{aligned}\frac{\partial}{\partial x}\frac{1}{\varepsilon}\frac{\partial H_z^{sc}}{\partial x} + \frac{\partial}{\partial y}\frac{1}{\varepsilon}\frac{\partial H_z^{sc}}{\partial y} + \omega^2 \varepsilon_0 \mu_0 \mu \, H_z^{sc} = \\ -\frac{\partial}{\partial x}\frac{1}{\varepsilon}\frac{\partial H_z^{in}}{\partial x} - \frac{\partial}{\partial y}\frac{1}{\varepsilon}\frac{\partial H_z^{in}}{\partial y} - \omega^2 \varepsilon_0 \mu_0 \mu \, H_z^{in}\end{aligned} \tag{4}$$

Introducing discrete grids for $x$ and $y$ with $n_x$ and $n_y$ grid points, respectively, and using the central difference approximation, we obtain a set of linear algebraic equations of the form

$$\mathbf{C}^{sc} \mathbf{h}_z^{sc} = \mathbf{s}^{in} , \qquad (5)$$

where $\mathbf{C}^{sc}$ is an N x N coefficient matrix, $N = n_x n_y$, $\mathbf{h}_z^{sc}$ is an N-vector corresponding to the unknown scattered field on the grid points. In the case of a plane wave incident field, the known, incident field $H_z^{in}$ is proportional to $\exp(i\mathbf{k}\cdot\mathbf{x})$, $\mathbf{k}$ = the desired incident wave vector and $\mathbf{s}^{in}$ is the discretized right-hand side of Eq. (4). For the case of point source(s), it suffices to take $\mathbf{s}^{in}$ to be a nonzero value at the source point(s) and zero elsewhere. Actually, the parts of $\mathbf{C}^{sc}$ that correspond to edges of the computational domain must also involve absorbing boundary conditions if the system is non-periodic. We employ a uniaxial perfectly matched layer (UPML) for this purpose [5].

The problem is to solve for the complex-valued vector, $\mathbf{h}_z^{sc}$, given the complex-valued matrix and vector, $\mathbf{C}^{sc}$ and $\mathbf{s}^{in}$. Note that $\mathbf{C}^{sc}$ is complex-valued because of the UPML absorbing boundary conditions *and* the fact that it is often the case that realistic materials will have complex-valued $\varepsilon$ and $\mu$, with the corresponding imaginary parts representing loss. We are also interested in hypothetical materials with no loss and negative values of $\varepsilon$ and/or $\mu$ in regions of space. In such cases we still find it necessary to introduce a small imaginary part to $\varepsilon$ or $\mu$ in order to achieve a stable solution. However, we can set these loss terms to be very small and thus approach closely the limit of purely real-valued $\varepsilon$ and $\mu$. Our tests show, for a LHM slab in air, that $\varepsilon = \mu = -1 + 10^{-8} i$ is still stable, and takes the same computer time as for larger loss cases.

We implemented this 2-D FDFD approach in MATLAB®, without attention to optimization of the complex linear equation solution of Eq. (5). Nonetheless a typical calculation with system dimensions as described in Sec. 3, with $\Delta x = \Delta y = 0.5$ nm resolution, requires just about five minutes of CPU time on a 1.4 GHz Apple G4 computer.

While developed in relation to the TM case and the Helmholtz equation for $H_z$, similar considerations apply to the TE case and $E_z$. Of course, for a true LHM with $\varepsilon = \mu = -1$, say, Eqs. (1) and (2) are of the same form and represent physically equivalent situations. We are currently working on a 3-D implementation of our approach.

**3. Subwavelength lensing effect for an *n* = -1 slab**

To illustrate the above FDFD method, a LHM slab is studied. The parameters of the lens (Fig. 1) are: $\lambda = 100$ nm, $d = 20$ nm, and $\varepsilon = \mu = -1 + 10^{-8}i$. In order to get good subwavelength resolution impedance matching conditions, $|\varepsilon| \approx \varepsilon_{1,2}$, $|\mu| \approx \mu_{1,2}$, have to be met at the surfaces of the LHM slab [4]. We thus set $\varepsilon_1 = \varepsilon_1 = \mu_1 = \mu_2 = 1$ (air) for the material surrounding the slab. A point source is put $s = 10$ nm below the slab. Grid spacings are taken to be $0.005\lambda = 0.5$ nm. Fig. 2 (a) shows the distributions of the electric field on the image plane. The evanescent components of the source are seen to excite a surface mode of the slab, which is possible because of the loss we have introduced into the material parameters. While detrimental to achieving perfect imaging [9], it is an unavoidable aspect of many numerical simulation techniques [7] and actual experimental problems with natural loss. Unlike previous FDTD results [7,8], however, the surface mode is not as extended along the surface. The reason might be that, in the frequency domain, we get a steady state for a single frequency, the parallel component of the wave vector for the evanescent waves decay exponentially, and cannot propagate to a long distance. This kind of limited surface mode extension/coupling into is enough to give the subwavelength lensing effects. In Fig. 2 (b), we plot the electric field intensity along the object and the image planes. From Fig. 2(b) we see the resolution (full width at half maximum) is around $0.05\lambda \approx 5$ nm, i.e. ten times better than the diffraction-limited value ($\lambda/2 \approx 50$ nm). Note that the image line result has not been shifted relative to the object line profile, i.e. there is almost complete recovery of the original signal intensity magnitude. This calculation is consistent with the resolution improvement factor predicted by the analysis of Ref. [9], $-\{(2\pi d)^{-1}\ln|(\mu+1)/2|\}\lambda \approx 7$. Previous calculations on somewhat differently configured LHM systems have not yielded this good a resolution [6-8, 11, 12]. Note that the resolution factor decreases only logarithmically with $|1+\mu|$, which means it is very difficult in finite precision arithmetic to achieve better resolution for the present problem.

We have also examined the case of two point sources along the object line, separated by just 10 nm. Fig. 3 shows the object and image line intensities that result. All other details are as for the previous single source example. Note that the two sources can readily be separated in the image plane intensity pattern.

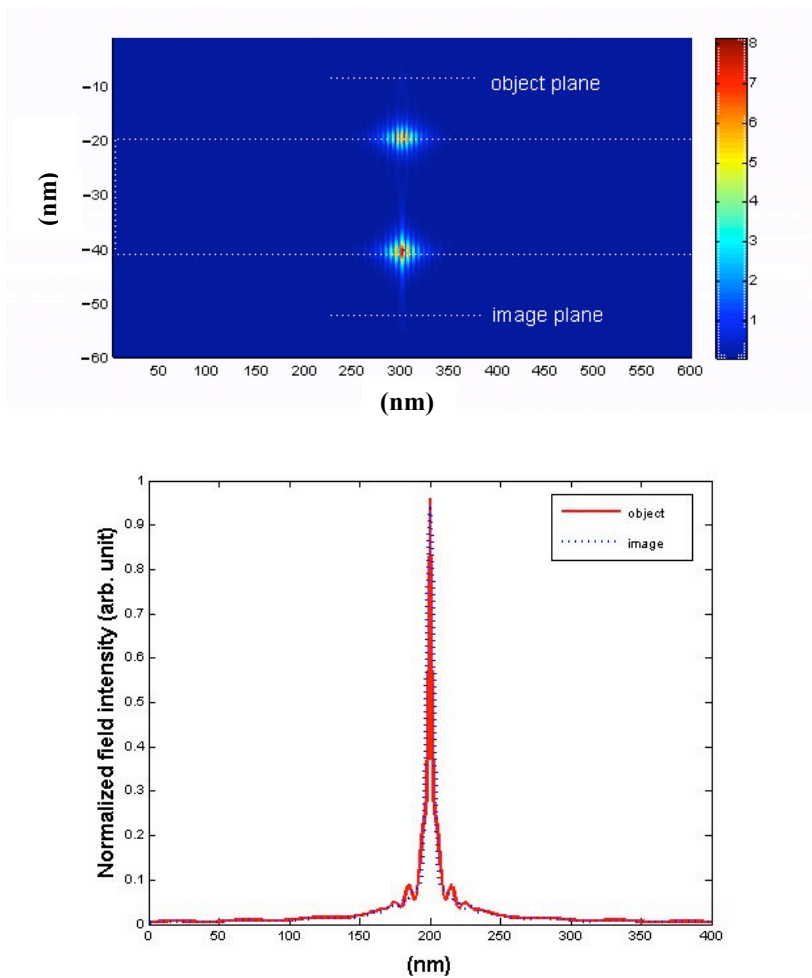

Fig. 2. Top panel: full spatial variation of the electric field intensity resulting from the point source/LHM problem discussed in the text. Bottom panel: intensity profiles along the object (solid red) and the image (dotted blue) planes.

## 4. Subwavelength resolution for a silver slab

In the electrostatic limit, the electric and magnetic response of materials can be decoupled. Pendry argued, for the TM mode in this case, that only the negative permittivity condition needs to be considered [4]. This makes noble metals such as silver good candidates for optical lensing [15, 16]. Here we use the FDFD method to simulate the experimental setup recently used for subwavelength lensing with a silver slab [16]. The object is an array of slits in chrome (assumed to be a perfect electrical conductor), each separated by 120 nm and slit widths of 60 nm, as indicated in Fig. 4. The object is placed on a 30 nm buffer material, which in turn rests on a 35 nm silver slab. On the other side of the silver slab is a photoresist (PR) material. The buffer and photoresist surrounding the silver slab are chosen to have permittivity +2.4. Accordingly, the working wavelength is chosen to be $\lambda$ = 365 nm, at which, the permittivity of silver is $\varepsilon$ = -2.4 + 0.25$i$ [16].

First, we consider a control experiment in which the silver slab is replaced by a 35 nm buffer material. A plane wave is assumed to be incident from the top of slits. Fig. 4(a) gives the simulation result. Due to the much smaller slit structure (60 nm) compared with wavelength (365 nm), there is no clear image of the slits and thus no evidence of subwavelength lensing effect. This is the conventional diffraction limit. Fig. 4(b) shows the simulation result with the silver slab. Even though the wavelength is much larger (six times) than the slit structure, we still get an image of the slits near the other side of the silver slab. We should note that the resolution we are discussing concerns the horizontal or $x$ direction. There are not sufficient propagating waves generated to develop a vertical image. (This was also true in the LHM slab cases discussed earlier; see also Ref. [12].

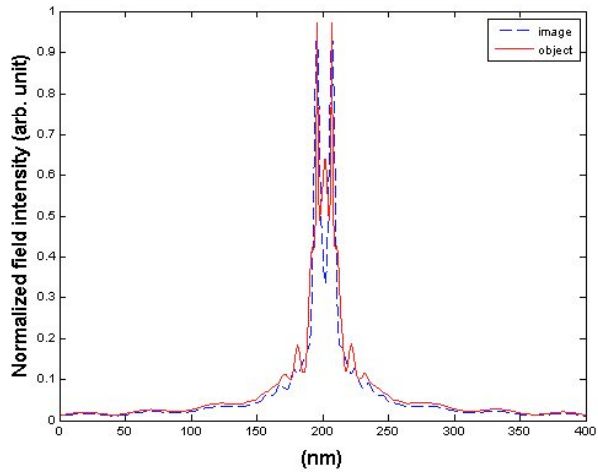

Fig. 3. Field intensity distributions along the object (solid red) and the image (dashed blue) planes for two point source/LHM case. The distance between the two point sources is 10 nm.

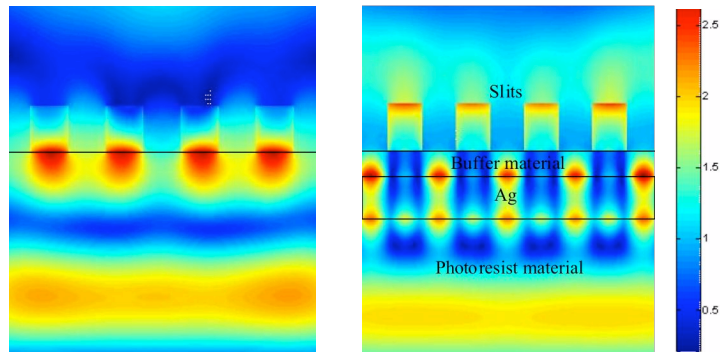

Fig. 4. Silver slab imaging. Left panel: field distribution for the control experiment discussed in the text. Right panel: the silver slab case. Solid lines show material boundaries.

## 6. Concluding Remarks

The finite-difference frequency-domain (FDFD) method, with the introduction of very small loss terms in the permittivity and permeability, provides a simple, direct means of simulating subwavelength imaging effects. The approach is very efficient in 2-D and can address a variety of experimentally relevant slab systems. We illustrated the approach by studying subwavelength imaging by a hypothetical, nanoscale left-handed material (LHM) slab with $n = -1$, as well as a simpler, more experimentally realizable silver slab [16]. Applications in 3-D are of particular interest in view of Pendry's recent proposal that a silver core-shell structure can exhibit lensing properties [17].

**Acknowledgments**

The work at Argonne National Laboratory was supported by the U. S. Department of Energy, Office of Basic Energy Sciences, Division of Chemical Sciences, Geosciences, and Biosciences under DOE contract W-31-109-ENG-38. Support at Northwestern is from US Department of Energy under grant No. DEFG02-03-ER15487, and from the NSF-supported Materials Research Science and Engineering Research Center (MRSEC). We are grateful to Drs. T. W. Lee and G. P. Wiederrecht for helpful comments.